# Do male leading authors retract more articles than female leading authors?

Er-Te Zheng[1], Hui-Zhen Fu[2], Mike Thelwall[1], Zhichao Fang[3,4*]

*Corresponding author: z.fang@cwts.leidenuniv.nl

1. School of Information, Journalism and Communication, The University of Sheffield, Sheffield, The UK.

2. Department of Information Resources Management, Zhejiang University, Hangzhou, China.

3. School of Information Resource Management, Renmin University of China, Beijing, China.

4. Centre for Science and Technology Studies (CWTS), Leiden University, Leiden, The Netherlands.

**Highlights**

- Male leading authors have a higher overall retraction rate than female leading authors.

- Gender disparities in retractions vary across retraction reasons and subject fields.

- Plagiarism and authorship issues are particularly more prevalent among male researchers.

**Abstract:** Scientific retractions reflect issues within the scientific record, arising from human error or misconduct. Although gender differences in retraction rates have been previously observed in various contexts, no comprehensive study has explored this issue across all fields of science. This study examines gender disparities in scientific misconduct or errors, specifically focusing on differences in retraction rates between male and female first authors in relation to their research productivity. Using a dataset comprising 11,622 retracted articles and 19,475,437 non-retracted articles from the Web of Science and Retraction Watch, we investigate gender differences in retraction rates from the perspectives of retraction reasons, subject fields, and countries. Our findings indicate that male first authors have higher retraction rates, particularly for scientific misconduct such as plagiarism, authorship disputes, ethical issues, duplication, and fabrication/falsification. No significant gender differences were found in retractions attributed to mistakes. Furthermore, male first authors experience significantly higher retraction rates in biomedical and health sciences, as well as in life and earth sciences, whereas female first authors have higher retraction rates in mathematics and computer science. Similar patterns are observed for corresponding authors. Understanding these gendered patterns of retraction may contribute to strategies aimed at reducing their prevalence.

## 1. Introduction

The reliability of scientific findings is intrinsically tied to the integrity of the research process, which requires researchers to present accurate and truthful representations of both the natural and social worlds (Bornmann, 2013). Research integrity has long been regarded not only as a fundamental principle of sound research practices but also as essential for maintaining public trust and confidence in science (Kreutzberg, 2004). However, numerous real-world examples reveal that researchers occasionally engage in misconduct that compromises the credibility of the scientific literature (Fanelli, 2009).

Scientific misconduct encompasses a wide range of unethical behaviors. As defined by Fanelli (2013), it involves "any omission or misrepresentation of the information necessary and sufficient to evaluate the validity and significance of research, at the level appropriate to the context in which the research is communicated". Misconduct can manifest in various forms, including data fabrication or falsification, plagiarism, and disputes over authorship (Bornmann, 2013). Such misbehavior has long been a serious concern for both the scientific community and governmental bodies worldwide. The primary consequence of these misbehaviors – problematic articles – can spread misinformation, waste resources, erode public trust in science, and in extreme cases, endanger public safety (Grieneisen & Zhang, 2012; Marcus, 2018).

To mitigate these issues, retraction serves as a vital mechanism for correcting the scientific record (COPE Council, 2019). Articles may be retracted due to various forms of misconduct, or alternatively, by authors laudably acknowledging errors or other problems in one of their studies (Nath et al., 2006). Over the past few decades, the number of retracted articles has steadily increased (Van Noorden, 2023), probably due to the growing prevalence of scientific misconduct, increased data sharing, and more rigorous editorial practices aimed at addressing problematic publications (Brainard, 2018). While this rise in retractions has raised concerns about the overall integrity of science, many scientists and editors view it as a positive sign, indicating that the scientific community is becoming increasingly vigilant and responsive to misconduct (Fanelli, 2013b; Van Noorden, 2011). For scientometric researchers, retractions offer concrete data for tracking scientific misconduct or errors, providing evidence of the self-correcting nature of science.

*1.1 Scientometric research on retracted articles*

Previous research on article retractions has primarily focused on two key areas: the bibliometric characterization of retracted articles and the measurement of their impact. Although the proportion of retracted articles remains relatively low, their absolute number has increased in recent years (Steen, 2011; Van Noorden, 2023). Many scholars believe that the true prevalence of problematic articles far exceeds the number of reported retractions (Fanelli, 2013b). The rate of retractions varies significantly across scientific disciplines (Grieneisen & Zhang, 2012; Lu et al., 2013), countries (Amos, 2014; Khademizadeh et al., 2023), journals (Cokol et al., 2007; Fang & Casadevall, 2011), and reasons (Fang et al., 2012; Steen, 2011). These findings have spurred further investigation into the risk factors associated with retractions, such as academic culture, financial incentives for publication, and national policies on misconduct (Fanelli et al., 2015, 2019). Additional research has shown that gold open access associates with faster retractions, and that intensive social media attention can increase the likelihood of retraction for misconduct-related research (Shema et al., 2019; Zheng et al., 2024).

Regarding impact measurement, the retraction process generally leads to a marked and sustained decline in the annual number of citations received by retracted articles (Furman et al., 2012; Shuai et al., 2017). Despite this decline, citations and Mendeley readership often continue to grow even after an article has been retracted (Bar-Ilan & Halevi, 2018; Budd et al., 1998). Moreover, a significant portion of post-retraction citations remain positive, regardless of the reasons for retraction (Bar-Ilan & Halevi, 2017). Retracted articles also tend to attract substantial attention on social media platforms (Khan et al., 2022), often at levels surpassing the attention given to non-retracted articles (Dambanemuya et al., 2024; Peng et al., 2022; Serghiou et al., 2021). The sentiment of Twitter mentions related to retracted articles typically skews negative, both before and after retraction (Amiri et al., 2024), suggesting that social media may play a role in bringing attention to problematic publications (Haunschild & Bornmann, 2021).

*1.2 Gender disparities in article retractions*

Gender disparities in scientific research have been well-documented across various dimensions, including research outputs (Huang et al., 2020), citations (King et al., 2017), funding (Larivière et al., 2011), and peer review (Lerback & Hanson, 2017). Some studies have shown that female researchers tend to publish fewer articles than their male counterparts in many disciplines and countries (Aksnes et al., 2011; Fox, 2005), presumably due to part-time working and career gaps taken to manage the additional caregiving responsibilities that many have. On average, male researchers produce between 16.8% and 31.6% more research outputs than female researchers over a given period (Abramo et al., 2009; Huang et al., 2020; Larivière et al., 2011). Since academic productivity typically increases with career progression, and men are overrepresented in senior positions, the gender disparity in research outputs tends to widen over time (Bordons et al., 2003).

While extensive research has been conducted on gender differences in research outputs, less attention has been paid to gender disparities in article retractions. Some studies suggest that men are more frequently associated with retracted articles than women, though these findings are often based on small datasets. For example, Decullier & Maisonneuve (2023) analyzed 113 retracted articles and found that 63% were authored by men. Similarly, Fang et al. (2013), in a study of 228 individuals involved in misconduct, reported that 65% were male. However, some researchers caution that a higher number of male-authored retractions does not necessarily indicate that men are more prone to misconduct. Instead, this disparity may reflect men's overrepresentation in research outputs (Fanelli, 2013b; Kaatz et al., 2013).

Empirical studies on gender and retractions have produced mixed results. For instance,

Fanelli et al., (2015), in an analysis of 661 retracted articles and 1,181 matched non-retracted controls, found no significant gender differences in retractions. Conversely, other research suggests that female researchers may be more susceptible to certain types of misconduct, such as some types of image duplications in biomedical research (Fanelli et al., 2019). These conflicting findings underscore the need for larger and more comprehensive studies to better understand the gender dynamics of scientific misconduct (Decullier & Maisonneuve, 2023).

*1.3 Objectives of the study*

Examining gender differences in retraction rates is essential for determining whether the retraction process is biased, potentially affecting male and female researchers differently. If articles authored by one gender are retracted more frequently for similar mistakes or misconduct, it may indicate systemic issues that require corrective action. Such an analysis ensures the retraction process remains equitable and upholds the integrity of scientific research.

The study aims to investigate gender disparities in article retractions, with a particular focus on the relationships between gender and the incidence and reasons for scholarly article retractions. By analyzing a large dataset of retracted articles in relation to global research productivity by gender, the study seeks to provide insights into the role of gender in research integrity and the broader dynamics of scholarly publishing. The following research questions (RQs) guide the investigation:

- **RQ1**. Do male leading (i.e., first and corresponding) authors have a higher rate of article retractions compared to female leading authors, relative to their research productivity?

- **RQ2**. Do gender disparities in article retractions vary between publication years, retraction reasons, scientific disciplines, and countries?

## 2. Data and methods

*2.1 Dataset of retracted and non-retracted articles*

In January 2025, we compiled a dataset of retracted articles published between 2008 and 2023. This dataset was derived by selecting articles labeled as "retracted publication" from the *Science Citation Index-Expanded* (SCIE), the *Social Sciences Citation Index* (SSCI), and the *Arts & Humanities Citation Index* (A&HCI) within the *Web of Science* (WoS) database. After manual verification, the dataset included 21,976 retracted articles. Bibliometric information, such as article titles, publication venues, publication years, DOIs, and authorship details, was extracted from the in-house WoS database maintained by the Centre for Science and Technology Studies (CWTS) at Leiden University.

Detailed information regarding retractions, particularly the reasons for retraction, was collected from the *Retraction Watch* database using DOIs of retracted articles. The Retraction Watch database is widely recognized as the most comprehensive repository for retracted articles (Brainard, 2018).

To enable a comparison of gender disparities among authors of retracted articles within a global context, we also gathered data on all non-retracted articles (i.e., articles that had not been retracted during the data collection period) from the SCIE, SSCI, and A&HCI for the same time frame (2008-2023). This resulted in a control group of 25,387,474 non-retracted articles. Bibliometric information for these articles was similarly retrieved from the CWTS in-house WoS database.

## 2.2 Gender inference of authors

To determine the gender of authors, we used a customized version of the WoS database with integrated gender identification, hosted by the CWTS at Leiden University. This version supports SQL-based queries and incorporates an author disambiguation algorithm (Caron & Eck, 2014), which is particularly well-suited for our analysis. Previous studies have demonstrated that this algorithm achieves a high level of precision (97%), although it may fail to capture some articles (recall = 90% to 91%) (Andersen & Nielsen, 2018; Caron & Eck, 2014), outperforming other similar author disambiguation approaches (Tekles & Bornmann, 2020).

Based on the author disambiguation process, the gender database assigns gender in a binary manner (i.e., male or female), inferred using a combination of three sources: Gender API, Genderize.io, and Gender Guesser. The gender inference primarily relies on the author's first name and the country of affiliation (Boekhout et al., 2021).[1] Gender inference is only made when the accuracy of the identification is reported to be at least 90%. Consequently, the gender of some authors may be labeled as "unknown" in cases of gender-ambiguous names or insufficient data regarding a particular name in a specific country (Madsen et al., 2022). This method of gender inference has been widely applied in gender-related studies (Andersen, 2023; Kozlowski et al., 2022; Madsen et al., 2022).

---

[1] See more information about the gender estimation applied also in the Leiden Ranking at: https://www.leidenranking.com/information/indicators.

For this study, we mainly focused on inferring the gender of the first author for each article, as the first author is typically responsible for the primary contribution to the research and is more likely to be held accountable for scientific misconduct (Decullier & Maisonneuve, 2023; Hussinger & Pellens, 2019; Larivière et al., 2016). We excluded articles where the first author's gender could not be determined. As shown in Table 1, this process resulted in a dataset of 11,622 retracted articles (52.9% of the total retracted articles) and 19,475,437 non-retracted articles (76.7% of the total non-retracted articles) for further analysis.

**Table 1**. Statistics of articles published (2008-2023) with inferred first author's gender.

| Article type | Number of articles with inferred gender | | Total articles with inferred gender (%) |
|---|---|---|---|
| | Male (%) | Female (%) | |
| Retracted articles | 8,088 (36.8%) | 3,534 (16.1%) | 11,622 (52.9%) |
| Non-retracted articles | 12,669,453 (49.9%) | 6,805,984 (26.8%) | 19,475,437 (76.7%) |

As with many previous studies, missing gender data remains an unavoidable challenge in gender-related research (see the statistics in Table A1 in the Appendix). Given that the corresponding author is responsible for the integrity of the manuscript content in some fields (Birnbaum et al., 2023), we conducted an additional analysis focusing on corresponding authors to further strengthen our findings and validate the results based on first-author data (see Section 3.6 and the Supplementary Materials).

*2.3 Classification of retraction reasons*

Retraction reasons are generally classified into two broad categories: misconduct and error (Feng et al., 2020; Resnik & Stewart Jr., 2012). Within these categories, more specific classifications have emerged, such as research misconduct, honest error, and publishing misconduct (Al-Hidabi & Teh, 2019; Hwang et al., 2023). Additional subcategories include issues like non-replicable findings, redundant publication, and unstated or unclear reasons (Wager & Williams, 2011). Furthermore, specific forms of misconduct, such as plagiarism and falsification, are often classified separately (Al-

Hidabi & Teh, 2019; Xu & Hu, 2022).

This study builds upon the classification frameworks for retraction reasons proposed by Tang et al. (2020) and Zhang et al. (2020). Drawing on their approaches, we refined and extended the classification based on the characteristics of our dataset and the scope of the research, resulting in nine distinct categories of retraction reasons, as outlined in Table A2 in the Appendix. It is important to note that the total number of retracted articles classified by different reasons exceeds the total number of retracted articles in our dataset, as 13.7% of the articles were retracted for multiple reasons. In our dataset, the most common retraction reason is "mistakes", followed by "fabrication/falsification" and "duplication". The average number of retraction reasons for articles retracted for multiple reasons is 2.10.

*2.4 The Leiden Ranking classification of subject fields*

For the disciplinary analysis, we adopted the Leiden Ranking classification (hereafter referred to as the LR classification, https://www.leidenranking.com/information/fields) to assign each retracted and non-retracted article to a specific subject field. The LR classification is a document-level system developed by Waltman and van Eck (2012). It organizes WoS-indexed publications – covering document types such as articles, reviews, letters, and proceedings papers – based on their direct citation relationships and clusters them into over 4,000 research areas using the Leiden algorithm (Traag et al., 2019). Each research area is then algorithmically linked to one of five primary subject fields (Table 2). The key advantage of the LR classification is its ability to assign each article to a specific subject field, which is particularly useful for articles published in multidisciplinary journals. It has been widely adopted in previous research related to disciplinary analysis (Didegah & Thelwall, 2018; Zahedi & Van Eck, 2018; L. Zhang et al., 2023). Table 2 lists the number of retracted and non-retracted articles in each subject field, along with the proportion of retracted articles relative to the total number of articles in each field.

**Table 2**. Distribution of retracted and non-retracted articles across five subject fields.

| Subject field | Number of retracted articles | Number of non-retracted articles | Proportion of retracted articles |
|---|---|---|---|
| Biomedical and health sciences | 6,196 | 7,753,324 | 7.99‰ |
| Physical sciences and | 1,044 | 3,005,331 | 3.47‰ |
| engineering | | | |
| Life and earth sciences | 1,551 | 1,753,800 | 8.84‰ |
| Mathematics and computer science | 2,143 | 5,039,688 | 4.25‰ |
| Social sciences and humanities | 688 | 1,923,294 | 3.58‰ |

*2.5 Indicators*

This study employs two indicators for the analysis: the *retraction rate* (RR), which measures the frequency of retractions for each gender, and the *male/female retraction ratio* (MFRR), which assesses gender disparities in retractions. The definitions and calculations of these indicators are outlined below:

- *Retraction rate* (RR): The RR for each gender is defined as the ratio of the number of retracted articles first-authored by a given gender to the total number of articles first-authored by that gender. The formula for calculating the RR is:

$$RR = \frac{NP_{retracted}}{NP_{all}}$$

where $NP_{retracted}$ denotes the number of retracted articles first-authored by a specific gender, $NP_{all}$ denotes the total number of articles first-authored by that gender, including both retracted and non-retracted articles. Confidence intervals for the RR were calculated using the Wilson Score interval.

- *Male/female retraction ratio* (MFRR): The MFRR is defined as the ratio of the retraction rate of male researchers to that of female researchers. The formula is:

$$MFRR = \frac{RR_{male}}{RR_{female}}$$

where $RR_{male}$ denotes the retraction rate for male researchers, and $RR_{female}$ denotes the retraction rate for female researchers within the same set of articles. Previous research often compares gender disparities by examining the absolute number of retracted articles. However, this approach does not account for differences in the total number of publications between male and female researchers. The MFRR offers a more objective comparison of retraction rates between genders. An MFRR value greater than 1 indicates a higher retraction rate for men compared to women, whereas a value less than 1 suggests a higher retraction rate for women compared to men. When the MFRR equals 1, it signifies identical retraction rates between genders. Confidence intervals for the MFRR were calculated using the Katz method (Katz et al., 1978).

## 3. Results

Based on our dataset, the overall retraction rate (RR) for male first authors between 2008 and 2023 is 6.38 per ten thousand articles published (‰₀), while for females, it is 5.19‰₀. This yields an MFRR of 1.23 (95% CI: 1.18, 1.28), indicating that male first authors generally have a higher retraction rate than females. In the following subsections, we investigate gender disparities in retractions across four dimensions: temporal trends, retraction reasons, disciplinary variations, and countries, along with a cross-analysis of disciplines and retraction reasons.

### 3.1 Temporal trends in retractions by gender

By aggregating articles by publication year, Figure 1 illustrates the temporal trends in gender disparities in retractions. Male first authors have consistently higher numbers of retractions (Figure 1a) and, in most years, higher retraction rates (Figure 1b) compared to their female counterparts. The gender disparities are further confirmed by the MFRR values (Figure 1c), which fluctuate between 1.0 and 1.5, remaining significantly greater than 1 during most of the examined period. This suggests that male first authors are proportionally associated with more retractions than female first authors.

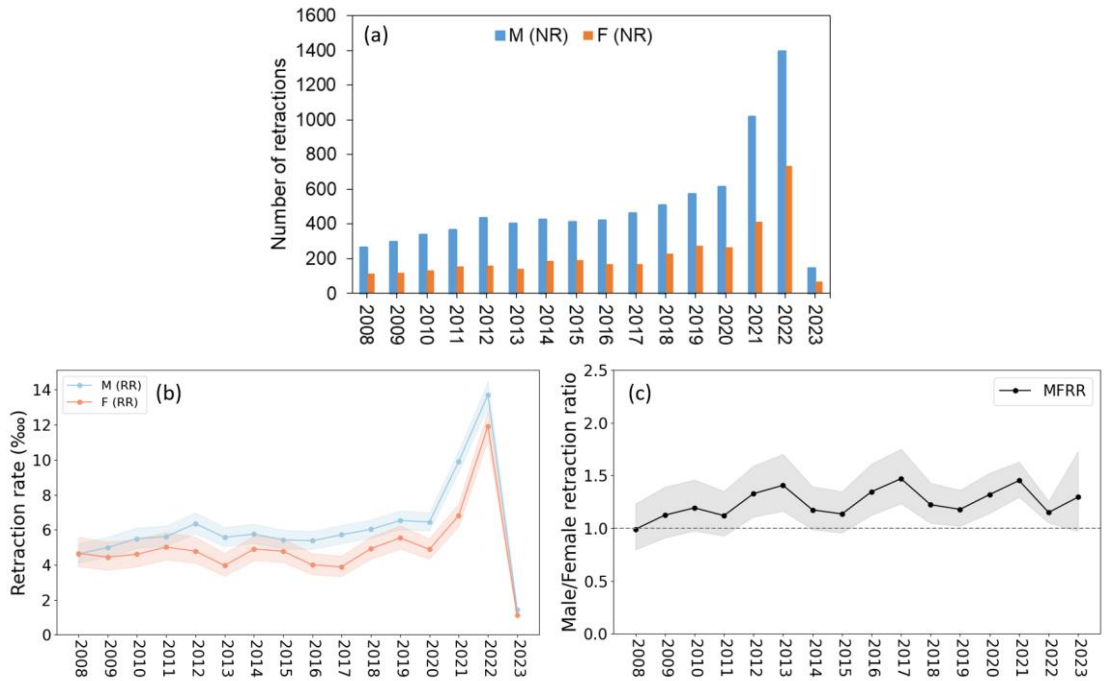

**Figure 1**. (a) Annual number of retractions (NR) by gender; (b) Annual retraction rate (RR) by gender with 95% confidence intervals; (c) MFRR temporal trend with 95% confidence intervals.

*3.2 Retraction reasons by gender*

For nearly all retraction reasons, whether a single reason or multiple reasons, male first authors have significantly higher retraction rates (Figure 2). The only exception is for "mistakes", where no significant gender difference is observed. Specifically, male first authors have much higher retraction rates for "plagiarism" (MFRR: 1.99) and "authorship issues" (MFRR: 1.73) compared to female first authors, although the overall retraction rates for these two reasons are relatively low. For retractions related to "duplication", "fabrication/falsification", and "ethical issues", male first authors also have higher retraction rates, though the gender differences are smaller.

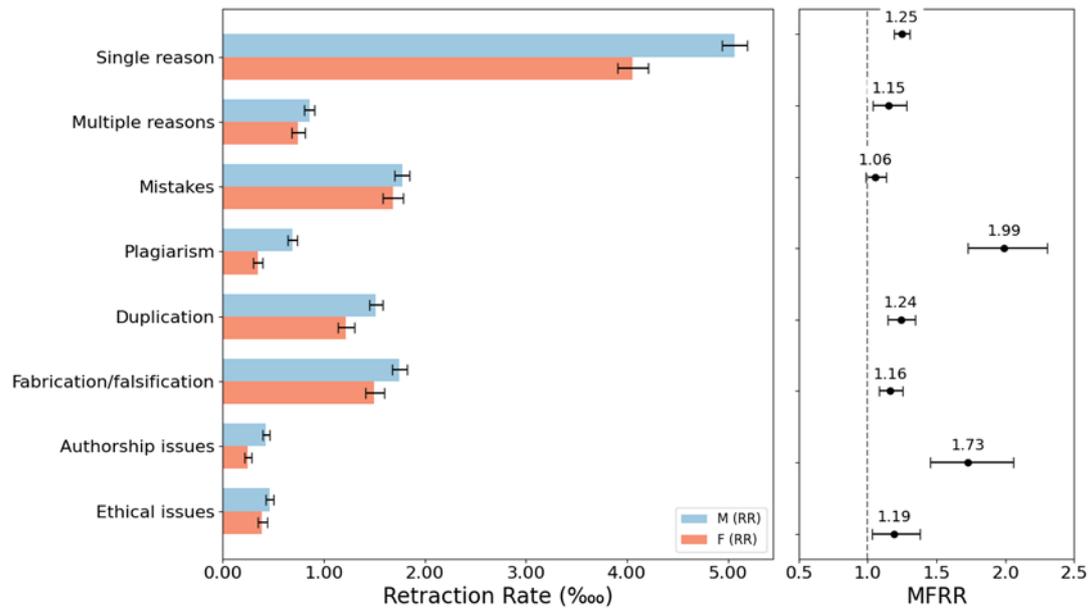

**Figure 2**. RR and MFRR by retraction reason for articles published between 2008 and 2023. Error bars indicate 95% confidence intervals.

Over the observed period, most categories of retraction reasons show an MFRR that remains close to 1 (Figure 3), indicating that, for the majority of retraction reasons (including multiple reasons), the annual retraction rates for male and female first authors do not differ substantially. However, for certain reasons, such as "plagiarism" and "duplication", the MFRR exceeds 1 more frequently, suggesting that the retraction rates of male first authors are higher than those of female first authors in these specific cases. Although the differences in retraction rates across years and retraction reasons are not large, the cumulative effect of all causes results in a higher overall male retraction rate. This explains why, in most years, the MFRR for "single reason" retractions exceeds 1.

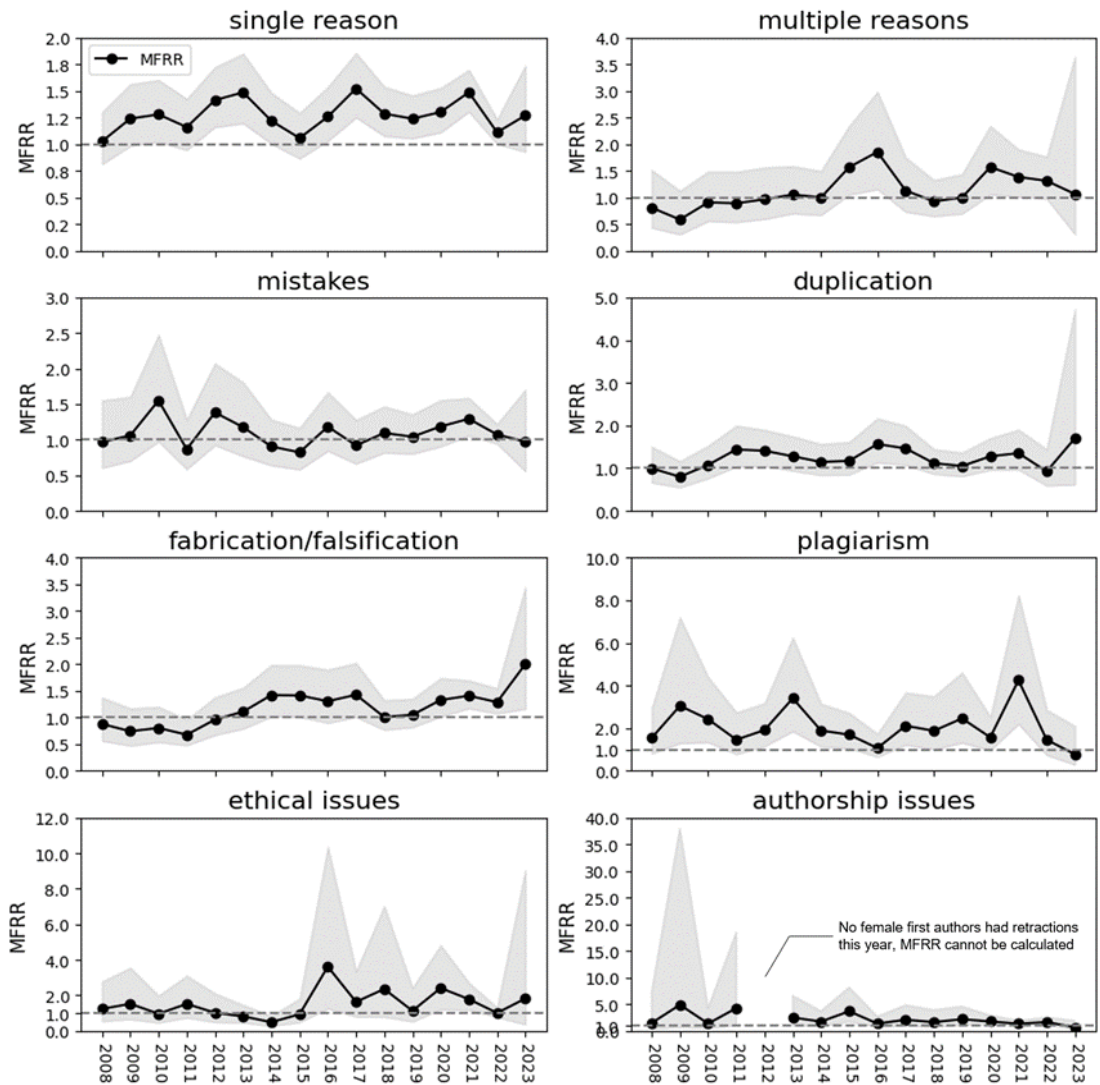

**Figure 3**. Annual MFRR by retraction reason for articles published between 2008 and 2023. Shaded areas indicate 95% confidence intervals.

### 3.3 Disciplinary variations of retractions by gender

In the fields of biomedical and health sciences, life and earth sciences, and physical sciences and engineering, male first authors have significantly higher retraction rates compared to their female counterparts. In contrast, within the mathematics and computer science field, female first authors have significantly higher retraction rates. In social sciences and humanities, no significant differences in retraction rates are observed between male and female first authors (Figure 4).

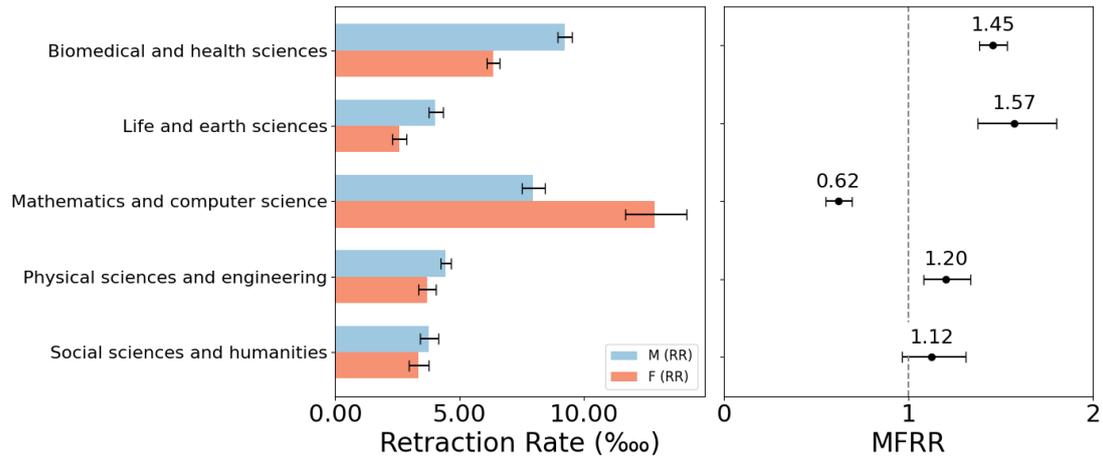

**Figure 4**. RR and MFRR across subject fields. Error bars indicate 95% confidence intervals.

### 3.4 Country variations of retractions by gender

We examined the ten countries with the highest number of articles retracted by first authors. In half of these countries, no significant differences in retraction rates are observed between male and female first authors (Figure 5). However, significant gender disparities are observed in certain countries: for Iran, Pakistan, and the United States, male first authors have higher retraction rates, whereas for Italy and China, female first authors have higher retraction rates. For most countries, the proportion of first authors with unidentified gender was relatively low. However, for China, many Chinese names cannot be easily classified by gender based on their English transliterations, which introduced potential challenges for accurate gender classification. As a result, the findings for China remain subject to further validation.

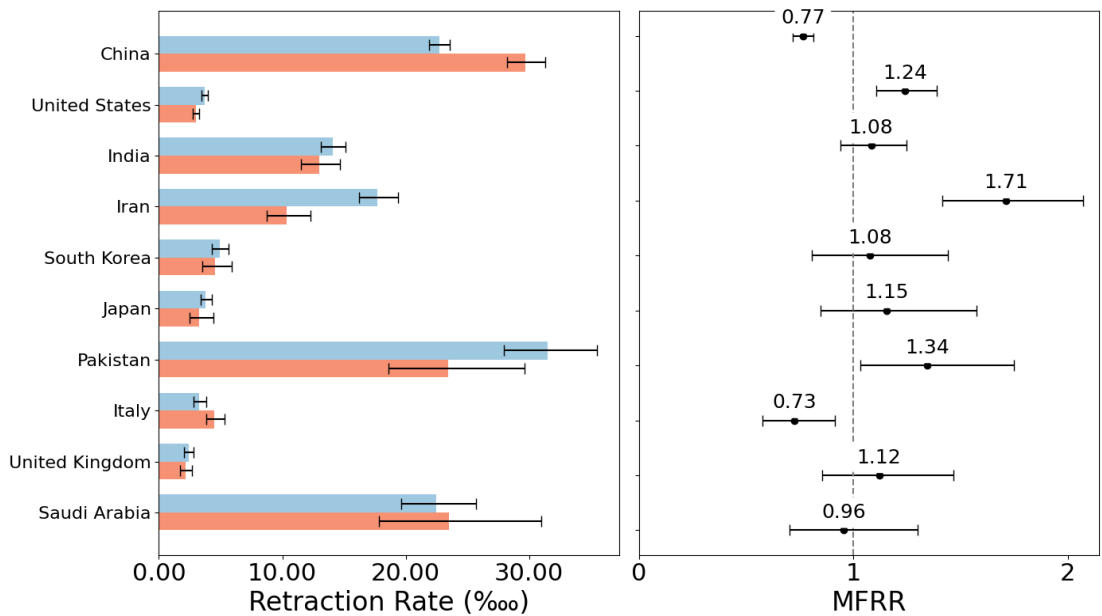

**Figure 5**. RR and MFRR across countries. Error bars indicate 95% confidence intervals.

*3.5 Cross-analysis of disciplines and retraction reasons*

To gain a deeper understanding of the varying retraction rates between disciplines, we conducted a cross-analysis of the MFRR, segmented by both subject fields and reasons for retraction (Figure 6). While male first authors generally have higher retraction rates across most categories, the MFRR values and confidence intervals show considerable variations, likely influenced by smaller sample sizes in certain categories.

In biomedical and health sciences and life and earth sciences, male first authors have higher retraction rates than their female counterparts for nearly all retraction reasons. In mathematics and computer science, female first authors have higher retraction rates for both single and multiple retraction reasons, as well as for mistakes, fabrication/falsification. No significant gender differences are observed for the other retraction reasons. In physical sciences and engineering, male first authors have higher retraction rates for single reason retractions, duplication, and plagiarism. Similarly, in social sciences and humanities, male first authors have higher retraction rates for duplication, plagiarism, and authorship issues.

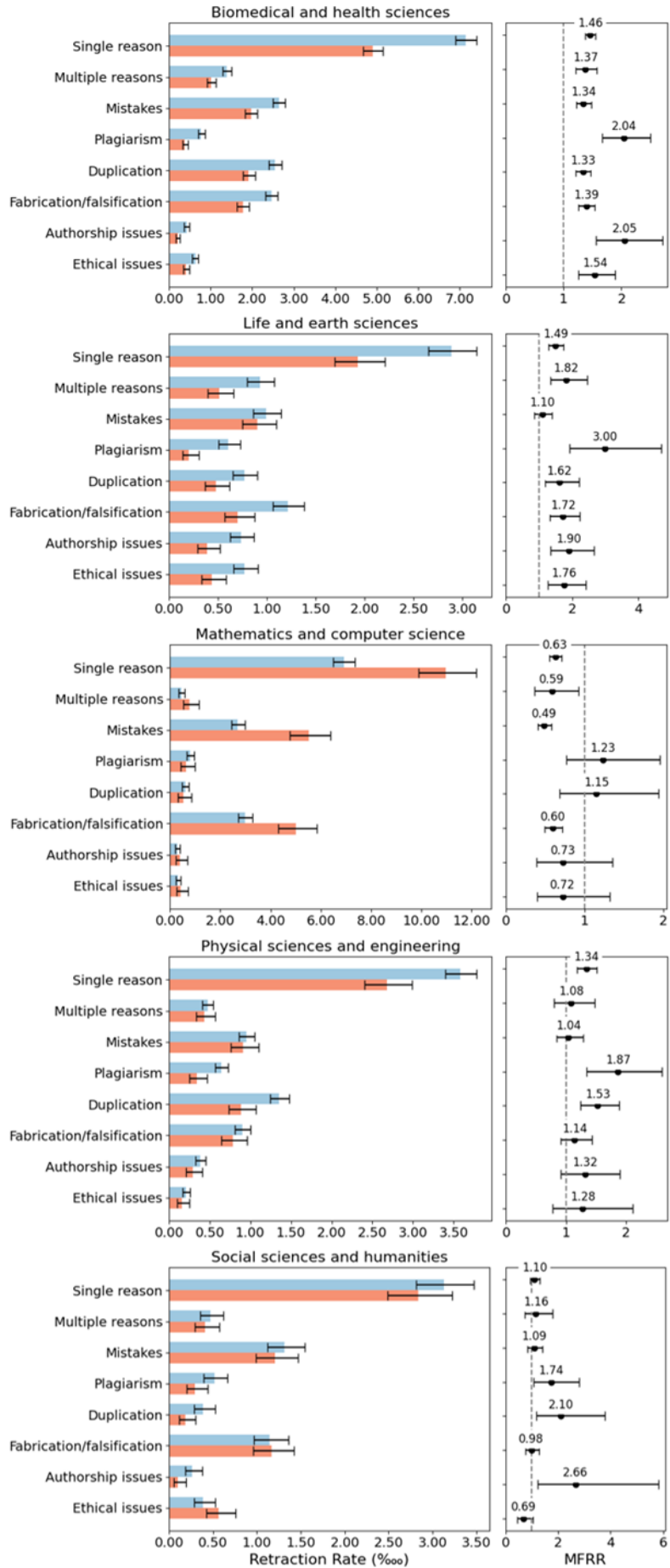

**Figure 6**. Cross-analysis of RR and MFRR by subject field and retraction reason. Error bars indicate 95% confidence intervals.

*3.6 Robustness check with corresponding authors*

Given the limitations in gender data availability, we performed additional analyses to assess the robustness of our findings. Specifically, we expanded our analysis to include corresponding authors, with detailed results provided in the Supplementary Materials. The key findings from this corresponding-author analysis are summarized as follows:

- Male corresponding authors have a higher overall retraction rate compared to female corresponding authors, with an MFRR of 1.20 (95% CI: 1.15, 1.25).

- For most retraction reasons – except for "mistakes" and "ethical issues" – male corresponding authors have higher retraction rates than female corresponding authors.

- In biomedical and health sciences and life and earth sciences, male corresponding authors have higher retraction rates. However, in mathematics and computer science, female corresponding authors have higher retraction rates. In physical sciences and engineering, no significant gender difference is found, while in social sciences and humanities, male corresponding authors show slightly higher retraction rates than their female counterparts.

- In half of the top ten countries with the highest number of retracted articles, no significant gender differences in the retraction rates of corresponding authors are observed. However, China has a higher retraction rate for female corresponding authors, while the United States, Iran, Pakistan, and Egypt have higher retraction rates for male corresponding authors.

Overall, the main conclusions regarding the retraction rates of male and female researchers remain consistent between the first-author and corresponding-author analyses. This consistency provides further support for the robustness of our findings.

## 4. Discussion

This study used a dataset of retracted articles, along with global research productivity data from 2008 to 2023, to explore gender disparities in article retractions across multiple dimensions.

*4.1. Higher retraction rates among male researchers*

Our findings, which reveal higher retraction rates for male researchers compared to female researchers, are consistent with previous studies that report a greater number of retracted articles authored by men (Decullier & Maisonneuve, 2023) and a higher incidence of male involvement in scientific misconduct cases (Fang et al., 2013). Importantly, the lack of significant gender differences in retractions due to "mistakes" suggests that the observed disparity is primarily driven by scientific misconduct.

One potential explanation for the higher retraction rates among male researchers could be their greater propensity for engaging in scientific misconduct. Psychological studies on gender differences have demonstrated that men are generally more prone to risk-taking behaviors in various contexts, such as in traffic situations (Konecni et al., 1976; Pawlowski et al., 2008) and financial decision-making (Charness & Gneezy, 2012). According to *Parental Investment Theory* (Trivers, 1972), this tendency may have evolutionary origins, as men are more inclined to engage in risky behaviors due to a belief in potential rewards and a reduced perception of risk, driven by their higher tolerance for both physical and social harm (Schumann & Ross, 2010). Consequently, male researchers may be more likely to commit misconduct, motivated by the anticipated benefits and a lower perceived risk of adverse consequences, as noted by Ameen et al. (1996).

Another plausible reason for the gender disparity in retraction rates is that male researchers may simply be more likely to be detected for misconduct than female researchers (Fang et al., 2013; Kaatz et al., 2013). Not all instances of misconduct are detected (Fanelli, 2009), particularly those involving subtle forms like falsification (Smith, 2000), which are more challenging to identify. It is possible that the greater visibility of male misconduct is a contributing factor to the observed higher retraction rates among male researchers.

*4.2. Varying gender disparities between retraction reasons and disciplines*

While male researchers overall have higher retraction rates, gender disparities vary across retraction reasons and subject fields. For most retraction reasons, whether caused by a single factor or multiple factors, male researchers tend to have higher retraction rates, particularly for issues such as plagiarism and authorship disputes. This finding contrasts with the study by Fanelli et al. (2019), which found no significant gender difference in image manipulation cases among first or last authors in *PLoS One*. This discrepancy between the studies may be attributed to differences in sample sizes, which could influence the statistical significance of such findings.

Disciplinary analysis further reveals that while male researchers have higher retraction rates across most subject fields, female researchers have higher retraction rates in mathematics and computer science. These disciplinary differences in retraction rates suggest that certain forms of scientific misconduct may be more prevalent or detectable in particular fields. Gender disparities in misconduct also appear to vary by discipline, with distinct dynamics emerging in different scientific communities. This issue, however, has not been extensively explored in prior research, making our findings a valuable addition to the literature.

*4.3 Implications*

Previous studies examining gender differences in retractions have often relied on the absolute number of retracted articles (Decullier & Maisonneuve, 2023; Fang et al., 2013). However, since male researchers typically publish significantly more articles than their female counterparts (Abramo et al., 2009; Huang et al., 2020; Larivière et al., 2011), a higher number of retracted articles among male researchers does not necessarily indicate a greater propensity for misconduct. In contrast, this study addresses this limitation by accounting for the total number of publications and using the retraction rate (RR) and MFRR as comparative metrics. These indicators offer a more accurate assessment of gender differences in retraction risks, providing a methodological advancement over previous approaches.

Our findings not only reveal gender disparities in retraction rates but also raise important questions about the underlying factors driving these differences. For instance, why do male and female researchers experience significantly different retraction risks across disciplines, countries, and retraction reasons? These variations may point to a deeper connection between academic culture, competitive pressures, gender role expectations, and research integrity. Understanding how these complex factors contribute to gender disparities in article retractions could provide valuable insights into the patterns of scientific misconduct. Future research could explore the impact of these variables to deepen our understanding of scientific integrity and the academic culture in which misconduct occurs.

The cross-disciplinary gender differences in article retractions highlight the potential role of academic cultures and evaluation standards, which may vary significantly between disciplines. These findings suggest that discipline-specific policies related to research integrity are essential. Our results could inform the development of more targeted interventions, including tailored evaluation mechanisms within specific fields. Moreover, the gender disparities observed across countries may reflect differences in ethical oversight, academic evaluation processes, and cultural expectations. These disparities indicate that some countries may have gender imbalances in areas such as research integrity training, publication ethics regulations, and academic reward systems, potentially making particular gender groups more vulnerable to research misconduct.

*4.4. Limitations*

This study has several limitations that should be acknowledged. First, we assessed gender disparities using the gender of the first and corresponding authors, as is common in previous studies (Decullier & Maisonneuve, 2023; Hussinger & Pellens, 2019; Larivière et al., 2016). However, this approach does not account for the possibility that other authors may have contributed to the misconduct leading to retraction. In collaborative research, the involvement of both male and female authors could influence the results. Future studies could benefit from using contributorship statements in articles or individual retraction statements to better attribute responsibility for retractions.

Second, this study focuses on gender disparities in article retractions in relation to global research productivity but does not consider individual-level factors beyond gender and discipline. To gain a deeper understanding of the factors contributing to gender disparities, future research should investigate individual-level characteristics such as the academic rank of authors, as well as the integrity polices of their affiliated institutions or countries. These variables could provide further insight into the structural and personal factors that influence retraction rates.

Third, the classification of retraction reasons can be multifaceted. While the classification used in this study – based on Retraction Watch data – captures the primary retraction reasons, some more fine-grained and less common reasons, such as funding-related issues, are not represented. Therefore, to ensure a sufficient number of cases within each category, we adopted a broader classification scheme.

Lastly, although we were able to successfully identify the gender of both first and corresponding authors for approximately 50% to 80% of the retracted and non-retracted articles, a substantial proportion of authors had undetermined gender. Notably, over 80% of the authors with unidentified gender were affiliated with institutions in China, underscoring the need for improved gender classification methods for Chinese names in future studies. We also acknowledge the presence of non-binary authors in academia. However, due to limitations in both data availability and algorithmic capability, we were unable to infer non-binary gender identities based on author names. These limitations highlight the need to complement bibliometric approaches with qualitative methods or case-based analyses in future gender-related misconduct research.

## 5. Conclusions

This study reveals that male leading authors generally have higher article retraction rates compared to female leading authors. However, these gender disparities in retraction rates vary across retraction reasons, subject fields, and countries. Specifically, male leading authors tend to have higher retraction rates for most types of misconduct, including plagiarism, authorship issues, ethical issues, duplication, and fabrication/falsification. Additionally, the gender gap in retraction rates differs by subject field. Male leading authors have significantly higher retraction rates in biomedical and health sciences, as well as life and earth sciences, whereas female leading authors have higher retraction rates in mathematics and computer science.

The gender differences observed in misconduct types – such as the higher rates of plagiarism and authorship issues among male researchers – suggest distinct behavioral dynamics between male and female researchers. Understanding these differences could help identify the underlying motivations and gender-specific factors influencing scientific misconduct. Further research is needed to explore these dynamics in more detail and to better understand how gender and academic culture intersect in the context of research integrity.

## Declaration of competing interests

The authors have no conflicts of interest to declare.

## Acknowledgements

This study is supported by the National Natural Science Foundation of China (No. 72304274) and the Soft Science Research Program of the Zhejiang Provincial Department of Science and Technology (No. 2021C35040). Er-Te Zheng is financially supported by the GTA scholarship from the School of Information, Journalism and Communication of the University of Sheffield. Mike Thelwall is supported by the Fundação Calouste Gulbenkian European Media and Information Fund (No. 187003). The authors thank the three anonymous reviewers for their valuable suggestions.

## Appendix

**Table A1**. Number of samples with unidentified gender data in previous scientometric studies.

| Previous research | Tool used | Sample size | Number of samples with unidentified gender (percentage) |
|---|---|---|---|
| Szymula & Simova (2023) | Genderize.io | 949,533 | 485,826 (51.2%) |
| Pinho-Gomes et al. (2023) | Gender API | 35,635 | 14,786 (41.5% for first authors); 15,222 (42.7% for last authors) |
| Ribeiro et al. (2023) | Genderize.io | 472 | 122 (25.8%) |
| Paul-Hus et al. (2015) | Methods developed by Larivière et al. (2013) | 1,028,382 | 332,196 (32.3%) |

**Table A2**. Categorization of retraction reasons.

| No. | Grouped reason | Reasons recorded by Retraction Watch | Number of retracted articles (proportion) | Illustrative examples from retraction notices |
|---|---|---|---|---|
| 1 | Mistakes | Error in image/data/text/results and/or conclusions/methods/materials (general)/cell lines/tissues/analyses; Concerns/issues about image/data/results/referencing/attributions; Contamination of reagents/materials (general)/cell lines/tissues; Unreliable data/image/results; Results not reproducible | 3,395 (29.2%) | "We received a report indicating the following: 'the Panel of Investigation found no evidence of intentional fabrication. However, due to flaws in the systematic review process, it is likely that there are additional errors in the publication. The Panel therefore cannot confirm that the results of the meta-analysis are wholly valid and recommend the paper be retracted.'" (*JAMA Internal Medicine* 2020, p.931) |
| 2 | Fabrication/falsification | Falsification/fabrication of results/image/data; Manipulation of results/images; Hoax publication; Paper mill; Sabotage of materials/methods | 3,238 (27.9%) | "Amgen requested the retraction as an outcome of an internal review where it was determined that one of the Amgen authors had manipulated specific experimental data presented in Figures 1 and 3" (*Cell Metabolism* 2015, p.532) |
| 3 | Duplication | Duplication of text/image/data/article; Euphemisms for duplication; Salami slicing | 2,753 (23.7%) | "The Editor-in-Chief has retracted this article because several images in this article appear to overlap with those of a previously published article by different authors" (*Oncogenesis* 2023, p.15) |
| 4 | Plagiarism | Plagiarism of text/image/data/article; Euphemisms for plagiarism | 1,107 (9.5%) | "The authors have plagiarized part of a paper that had already appeared in Biomedicine & Pharmacotherapy" (*Saudi Pharmaceutical Journal* 2020, p.639) |
| 5 | Ethical issues | Legal reasons/legal threats; Civil/criminal proceedings; Ethical violations; Lack of ethical approval; Informed/patient consent-none/retracted; Infringement of patient | 853 (7.3%) | "The IRB found that the study included data from between one and four therapy clients of the Maryland Psychotherapy Clinic and Research Laboratory (MPCRL) who either had not been asked to provide consent or had withdrawn consent for their data to be included in the research" (*Dreaming* 2023) |

| | | privacy; Lack of balance/bias issues; Conflict of interest; Copyright claims | | |
|---|---|---|---|---|
| 6 | Authorship issues | Forged authorship; Concerns/issues about authorship | 709 (6.1%) | "Recently, these authors plausibly changed their forenames/first names to repeat similar misconducts… No person of that name is or has been listed as a faculty member of that institution and that name is affiliated with no other published paper than this one. It is our understanding that this author does not exist" (*Marine Technology Society Journal* 2022) |
| 7 | Single reason | Articles with only one of retraction reasons No.1-6. | 9,178 (79.0%) | / |
| 8 | Multiple reasons | Articles with multiple retraction reasons from No.2-6. (Note that "mistakes" refers only to issues related to research mistakes and will not appear alongside other retraction reasons) | 1,595 (13.7%) | "This article has been retracted by Hindawi following an investigation undertaken by the publisher. This investigation has uncovered evidence of one or more of the following indicators of systematic manipulation of the publication process: (1) Discrepancies in scope; (2) Discrepancies in the description of the research reported; (3) Discrepancies between the availability of data and the research described; (4) Inappropriate citations; (5) Incoherent, meaningless and/or irrelevant content included in the article; (6) Peer-review manipulation" (*Contrast Media & Molecular Imaging* 2023) |
| 9 | Reasons uncategorizable or not available | | 849 (7.3%) | "This article has been withdrawn at the request of the authors and editor. The Publisher apologizes for any inconvenience this may cause." (*Neurobiology of Aging* 2015) |

# Supplementary Materials

We conducted the same analysis for corresponding authors as we did for first authors to assess whether the results would differ from the first-author analysis and would therefore depend on who was believed to be the leading author. As detailed in Table S1, this process resulted in a dataset of 12,110 retracted articles (55.1% of all retracted articles) and 20,059,091 non-retracted articles (79.0% of all non-retracted articles) with inferred gender of corresponding authors.

**Table S1**. Statistics of articles published (2008-2023) with inferred corresponding author's gender.

| Article type | Number of articles with inferred gender | | Total articles with |
|---|---|---|---|
| | Male (%) | Female (%) | inferred gender (%) |
| Retracted articles | 9,099 (41.4%) | 3,011 (13.7%) | 12,110 (55.1%) |
| Non-retracted articles | 14,349,975 (56.5%) | 5,709,116 (22.5%) | 20,059,091 (79.0%) |

Based on our dataset, the overall retraction rate for male corresponding authors between 2008 and 2023 is 6.34 per ten thousand articles published (‰), while for female corresponding authors, it is 5.27‰. This results in an MFRR of 1.20 (95% CI: 1.15, 1.25), indicating that male corresponding authors also have a higher retraction rate than their female counterparts.

## Temporal trends in retractions by gender

The annual number of retracted articles authored by male corresponding authors consistently exceeds that of female corresponding authors (Figure S1a), with a higher retraction rate observed for males (Figure S1b). The annual MFRR fluctuates between 1.0 and 1.5, except in 2023, when it surpasses the average level (Figure S1c).

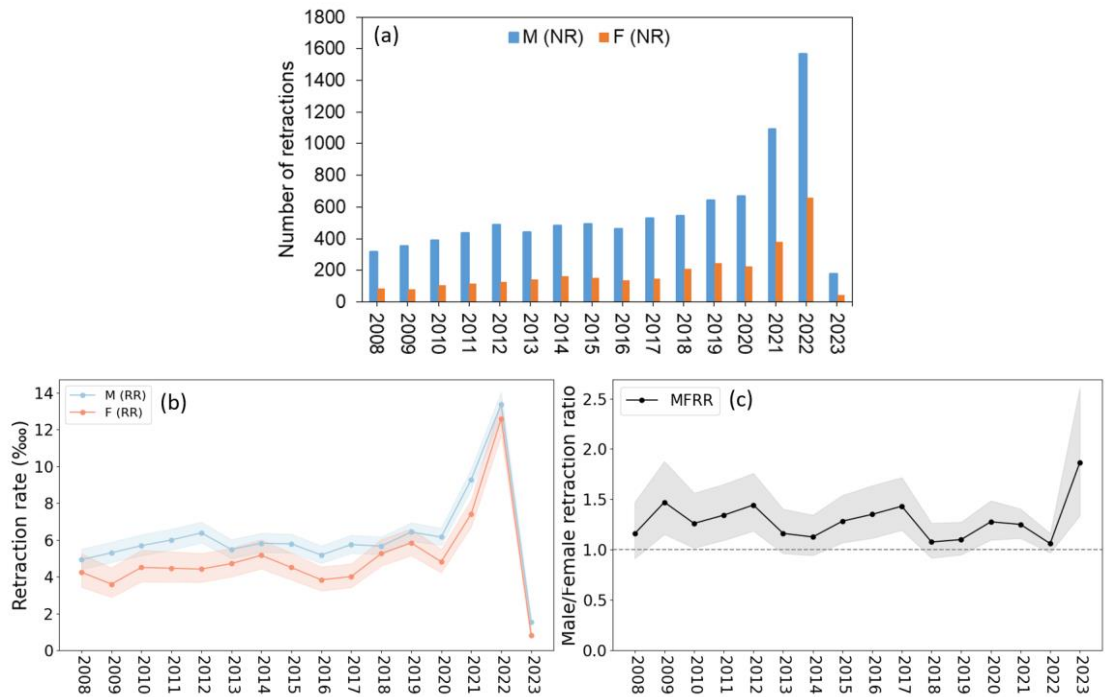

**Figure S1.** (a) Annual number of retractions (NR) identified by gender; (b) Annual retraction rate (RR) by gender with 95% confidence intervals; (c) MFRR temporal trend with 95% confidence intervals.

## Retraction reasons by gender

For nearly all retraction reasons, except for "mistakes" and "ethical issues", male corresponding authors have higher retraction rates (Figure S2). "Plagiarism" and "authorship issues" remain the most significant retraction reasons with the largest gender disparity, aligning with the findings for first authors.

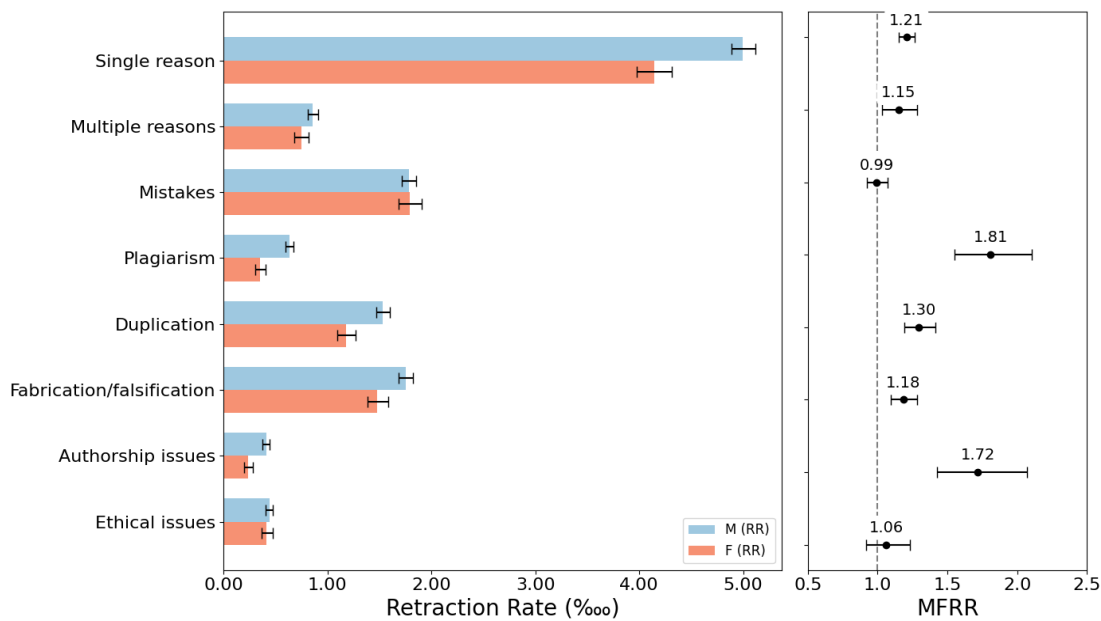

**Figure S2.** RR and MFRR by retraction reason for articles published between 2008 and 2023. Error bars indicate 95% confidence intervals.

While gender differences in retraction rates for most retraction reasons are not statistically significant in most years, there are more years in which male corresponding authors have higher retraction rates (Figure S3). Consequently, when aggregating results across all years, male corresponding authors show significantly higher retraction rates for the majority of retraction reasons.

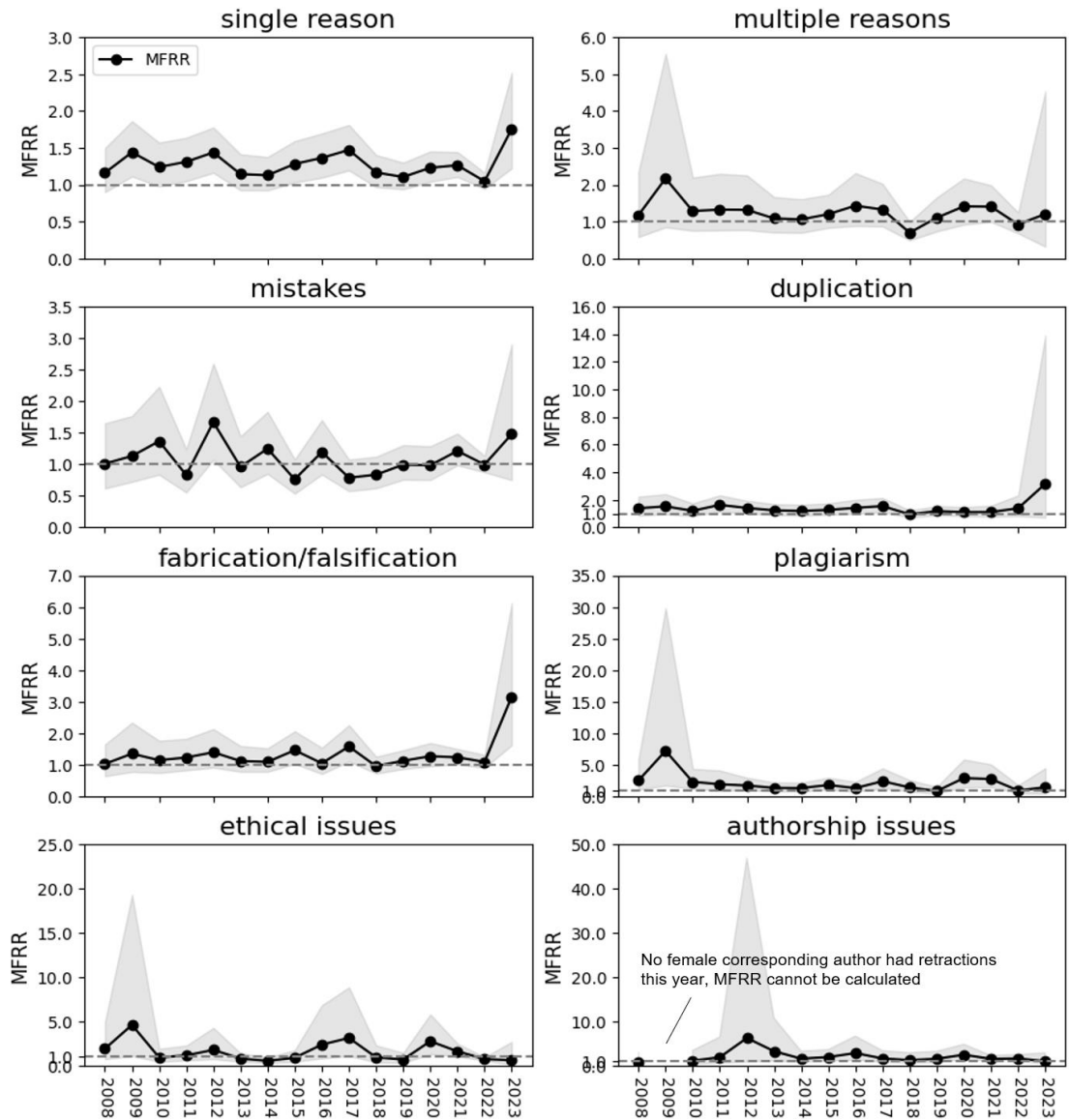

**Figure S3.** Annual MFRR by retraction reason for articles published between 2008 and 2023. Shaded areas indicate 95% confidence intervals.

**Disciplinary variations of retractions by gender**

In biomedical and health sciences, as well as life and earth sciences, male corresponding authors have higher retraction rates. However, in mathematics and computer science, female corresponding authors have higher retraction rates (Figure S4). This pattern is similar to the findings for first authors. The results for other disciplines show slight differences: in physical sciences and engineering, no significant gender difference is observed in retraction rates, while in social sciences and humanities, male corresponding authors have slightly higher retraction rates than females. Overall, the MFRR values for corresponding authors across various disciplines are highly consistent with those for first authors, reinforcing the robustness of the first-author analysis.

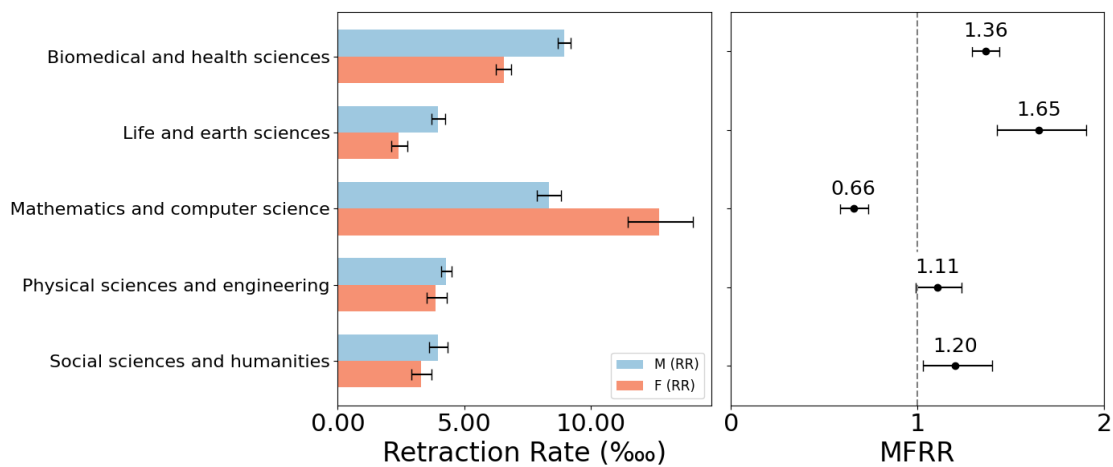

**Figure S4.** RR and MFRR across subject fields. Error bars indicate 95% confidence intervals.

**Country variations of retractions by gender**

Among the ten countries with the highest number of retracted articles by corresponding authors, the top nine countries are the same as for first authors, with Egypt replacing Saudi Arabia in the tenth position. In half of these countries, no significant gender difference is observed in the retraction rates of corresponding authors (Figure S5). China has a higher retraction rate for female corresponding authors, while the United States, Iran, Pakistan, and Egypt show higher retraction rates for male corresponding authors.

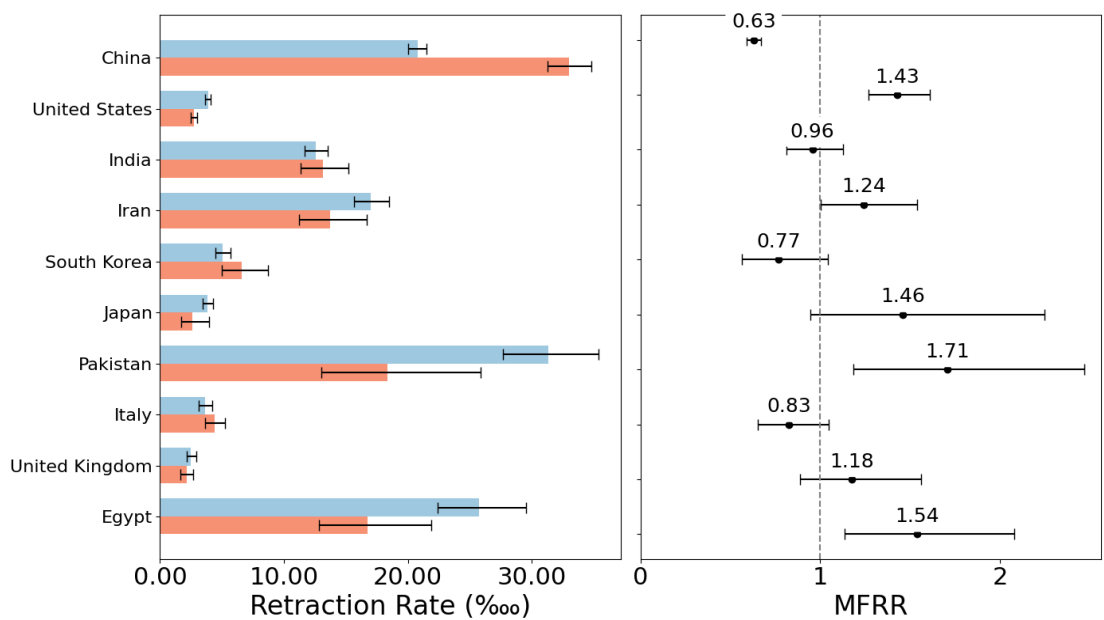

**Figure S5.** RR and MFRR across countries. Error bars indicate 95% confidence intervals.

**Cross-analysis of disciplines and retraction reasons**

In biomedical and health sciences and life and earth sciences, male corresponding authors have higher retraction rates for the majority of retraction reasons (Figure S6). In contrast, in mathematics and computer science, female corresponding authors have higher retraction rates for most retraction reasons. In physical sciences and engineering, as well as social sciences and humanities, there is no significant gender difference in retraction rates for most retraction reasons. However, in physical sciences and engineering, male corresponding authors have a higher retraction rate for "plagiarism", while in social sciences and humanities, male corresponding authors show higher retraction rates for "authorship issues", "duplication", and "plagiarism".

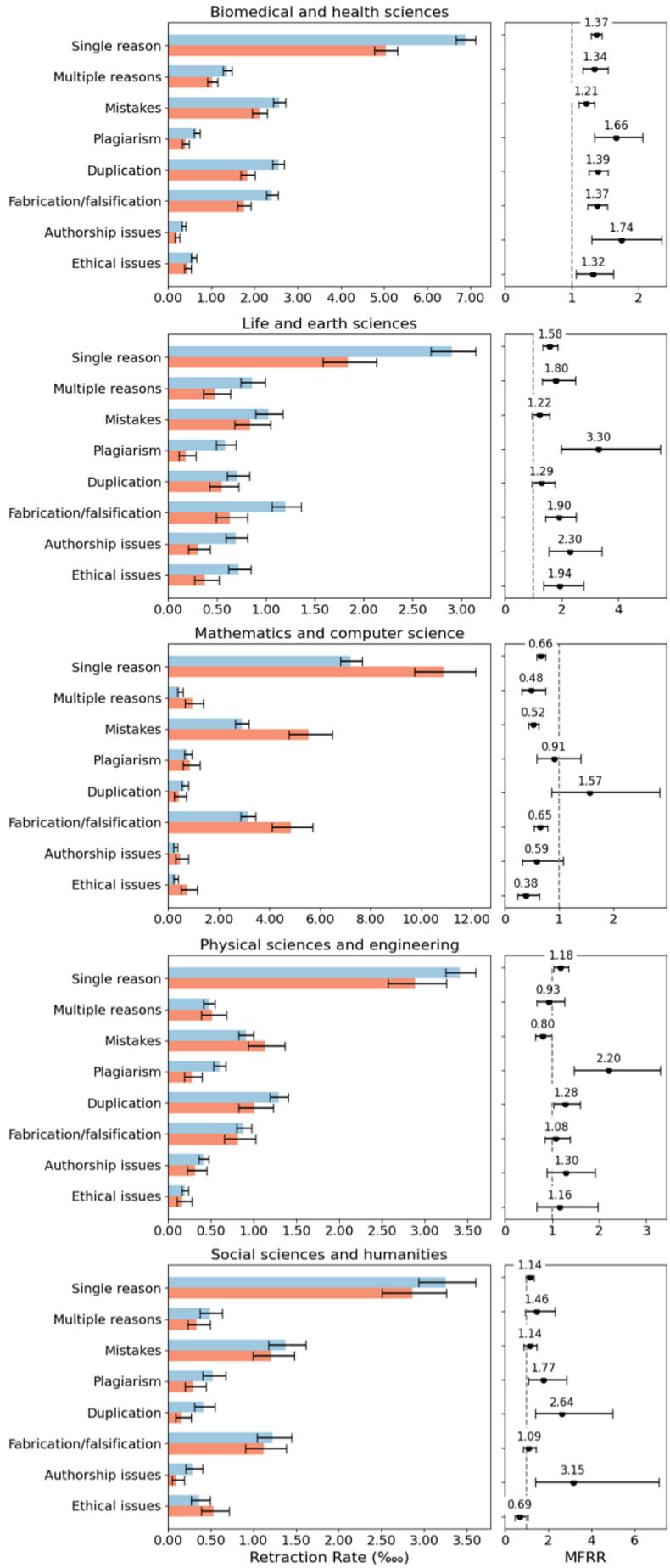

**Figure S6.** Cross-analysis of RR and MFRR by subject field and retraction reason. Error bars indicate 95% confidence intervals.